\documentclass[11pt]{article}
\usepackage{titlesec}
\usepackage{hyperref}

\titleclass{\subsubsubsection}{straight}[\subsection]

\usepackage[utf8]{inputenc}
\usepackage{amsmath,amsfonts,amssymb,stackengine,graphicx}
\author{meo}
\date{November 2025}

\usepackage[utf8]{inputenc}
\newif\iffigs\figstrue

\usepackage[title]{appendix}
\usepackage{epsfig,latexsym}
\usepackage{hyperref}
\usepackage{amsmath}
\usepackage{verbatim}
\usepackage{color}
\usepackage{mathrsfs}
\usepackage{slashed}
\usepackage{amssymb}

\DeclareMathAlphabet{\mathpzc}{OT1}{pzc}{m}{it}

 \csname
@addtoreset\endcsname{equation}{section}

\def\gz0{\gamma^{0}}

\def\beq{\begin{equation}}
\def\eeq{\end{equation}}
\def\bea{\begin{eqnarray}}
\def\eea{\end{eqnarray}}
\def\ba{\begin{array}}
\def\ea{\end{array}}
\def\bec{\begin{center}}
\def\ec{\end{center}}
\def\ba{\begin{align}}
\def\ena{\end{align}}

\def\12{\frac{1}{2}}

\newcounter{subsubsubsection}[subsubsection]
\renewcommand\thesubsubsubsection{\thesubsubsection.\arabic{subsubsubsection}}
\titleformat{\subsubsubsection}
  {\normalfont\normalsize\bfseries}{\thesubsubsubsection}{1em}{}
\titlespacing*{\subsubsubsection}
{0pt}{3.25ex plus 1ex minus .2ex}{1.5ex plus .2ex}

\makeatletter
\renewcommand\paragraph{\@startsection{paragraph}{5}{\z@}%
  {3.25ex \@plus1ex \@minus.2ex}%
  {-1em}%
  {\normalfont\normalsize\bfseries}}
\renewcommand\subparagraph{\@startsection{subparagraph}{6}{\parindent}%
  {3.25ex \@plus1ex \@minus .2ex}%
  {-1em}%
  {\normalfont\normalsize\bfseries}}
\def\toclevel@subsubsubsection{4}
\def\toclevel@paragraph{5}
\def\toclevel@paragraph{6}
\def\l@subsubsubsection{\@dottedtocline{4}{7em}{4em}}
\def\l@paragraph{\@dottedtocline{5}{10em}{5em}}
\def\l@subparagraph{\@dottedtocline{6}{14em}{6em}}
\makeatother

\begin{document}

\begin{flushright}
{\today}
\end{flushright}

\vspace{10pt}

\begin{center}

%%%%%%%%%%%%%%%%%%%%%%%%%%%%%%%%%%%%%%%%%%%%%%%%%%%%%%%%%%%%%%%%%%%%

{\Large\sc More on Pre--Inflationary Non Gaussianities}\\

%%%%%%%%%%%%%%%%%%%%%%%%%%%%%%%%%%%%%%%%%%%%%%%%%%%%%%%%%%%%%%%%%%%%

\vspace{25pt}
{\sc M.~Meo

{${}^c$\sl\small
Scuola Normale Superiore and INFN\\
Piazza dei Cavalieri, 7\\ 56126 Pisa \ ITALY \\
e-mail: {\small \it mario.meo@sns.it}}\vspace{10pt}
}

%%%%%%%%%%%%%%%%%%%%%%%%%%%%%%%%%%%%%%%%%%%%%%%
\vspace{40pt} {\sc\large Abstract}\end{center}
%%%%%%%%%%%%%%%%%%%%%%%%%%%%%%%%%%%%%%%%%%%%%%%
\noindent
I generalize the three--point amplitude of curvature perturbations in the climbing scenario inspired by ten--dimensional non--supersymmetric strings to a broader class of exponential potentials, under some assumptions on the smoothing effects of String Theory that favor a bounce Cosmology. The extension can encompass the $SO(16)\times SO(16)$ model, together with other scenarios related to supersymmetry breaking in String Theory. The $e$-fold ranges compatible with Planck data move toward lower values of $N$ for milder potentials and toward larger values for steeper ones.  I also compute the three-point amplitudes involving graviton modes in the same bounce scenario, showing in detail their lack of peculiar contributions from the turning point.

\setcounter{page}{1}

\pagebreak

\newpage
 \newpage
\baselineskip=20pt
%%%%%%%%%%%%%%%%%%%%%%%%%%%%%%%%%%
\section[Introduction]{\sc  Introduction and Summary}\label{sec:intro}
%%%%%%%%%%%%%%%%%%%%%%%%%%%%%%%%%%%%%
There are some indications in String Theory~\cite{stringtheory} that point to the breaking of supersymmetry~\cite{ssb_review} as a key ingredient in addressing the early stages of the Universe. String--theoretic realizations of supersymmetry breaking provide in fact clues of a peculiar ``climbing'' scenario for a pre--inflationary ``fast-roll to slow-roll'' transition~\cite{climbing} at weak string coupling, which resonates with the lack of power in the first few CMB multipoles~\cite{planck_lack}, and whose other possible implications for Cosmology deserve further investigations. This scenario is common to the three known tachyon--free non--supersymmetric string models, which are all characterized, in four--dimensional realizations, by the emergence of an exponential ``tadpole potential'' 
\beq
    V(\phi)\ = \ \frac{T}{2 \kappa^2}\ e^{\sqrt{6}\,\lambda \,\phi} \label{Vgamma_int}
\eeq
for the dilaton field $\phi$ that forbids the existence of a flat vacuum. In all three cases the climbing behavior is inevitable. $\lambda=1$, the ``critical'' value for its onset, for the two orientifold~\cite{orientifolds} models of~\cite{susy95} and of~\cite{usp32}, which is the simplest setting for ``brane supersymmetry breaking'' (for a short review, see ~\cite{bsb_rev}),  while $\lambda=\frac{5}{3}$, a ``super-critical'' value, for the heterotic model of~\cite{so1616}.

Spatially flat cosmological backgrounds display in fact a remarkable transition at $\lambda=1$ since, for this value or larger ones, the scalar field can no longer descend the potential when emerging from the initial singularity. Rather, it is compelled to emerge while climbing up, leading to a cosmological evolution that can unfold within the weak-coupling regime of String Theory and features a turning point in the scalar trajectory~\footnote{ The cosmological solutions, first found in~\cite{dm,russo}, revealed in~\cite{climbing} this peculiar transition.}. 
If this exponential tadpole term is accompanied by softer contributions — such as, for example, the Starobinsky potential that was a very significant step in the development of inflation~\cite{inflation} — the evolution can naturally trigger a slow-roll phase, as first discussed in~\cite{dkps}.

The pre-inflationary phase suppresses the power present at large wavelengths in the primordial power spectrum of curvature perturbations. In short--enough inflationary scenarios, the suppression can reduce the power present in the lowest multipoles of the CMB angular power spectrum~\cite{dkps}, in qualitative agreement with the observed power deficit present in those modes~\cite{planck_lack}. Following~\cite{dkps}, the effect can be parametrized by introducing a characteristic scale, $\Delta$, which controls the infrared suppression, before the power spectrum approaches the standard Chibisov--Mukhanov power--like behavior~\cite{cm}.  Interestingly, CMB observations provide some support for the presence of such a scale~\cite{gkmns}. In particular, with an enlarged mask around the Galactic plane corresponding to an open sky fraction of about $39\%$, which should be less sensitive to astrophysical backgrounds, the evidence for a nonzero $\Delta$ reaches the $3\sigma$ level~\cite{gkmns}. The inferred value is comparable to the present cosmic horizon and can be translated into an energy scale 
\beq
\Delta_{\text{inf}} \,\simeq\, 2 \times 10^{12}\, e^{N-60}\,\mathrm{GeV}
\eeq
at the onset of an inflationary phase of $N$ $e$-folds.

In~\cite{ms} we investigated the imprints of a pre--inflationary climbing phase on the bispectrum of curvature perturbations. The presence of a turning point in the scalar evolution, where the slow--roll parameter $\epsilon$ vanishes, introduces some subtleties in the formulation in terms of the Mukhanov--Sasaki variable. However, curvature perturbations remain well defined once the computation is consistently performed within the Schwinger--Keldysh formalism, and \emph{finite} contributions of a distinctive nature emerge from the vicinity of the turning point.
The analysis relied on some assumptions concerning the resolution of the initial singularity in a full string treatment, which was argued to have a smoothing effect, resulting in a bounce cosmology along the lines of the pre--Big Bang proposal of~\cite{gasp_ven}. It was found that the scale $\Delta$ induces characteristic oscillations in the nonlinearity parameter $f_{NL}$ for the equilateral configuration, and more importantly that the turning--point contribution selects a narrow window around $N \simeq 63$ for the inflationary $e$--folds, where the resulting non--Gaussianities can be sizable and yet compatible with current \textit{Planck} bounds~\cite{Planck,forecasts}, so that the effect could be tested in future experiments.

The analysis carried out in~\cite{ms} dealt with the special case of the potential in eq.~\eqref{Vgamma_int} with $\lambda=1$, which corresponds to the orientifold models of~\cite{susy95} and~\cite{usp32}. The microscopic dynamics near the turning point is somewhat simpler in that case, and moreover the analysis was confined to the three--point amplitude of scalar curvature perturbations, which were argued to be the only ones where the turning point would introduce a peculiar contribution. 
The purpose of the present work is to extend and complete the analysis, within similar assumptions on the smoothing effects of String Theory. 

After a brief review of the setup underlying the results obtained in~\cite{ms}, which will be presented in Section~\ref{sec:summary}, in Section~\ref{sec:lambdanotone} we shall address the three--point function of curvature perturbations for a generic exponential potential, characterized by an arbitrary exponent $\lambda$. As we shall see, the non--analytic behaviors manifested by the background solutions when approaching the critical limit $\lambda \to 1$ from the two regions with $\lambda<1$ and $\lambda>1$ will disappear in the final contributions, and the two sides will join smoothly to the results obtained in~\cite{ms} for $\lambda=1$. The result will also manifest an approximate scaling behavior with $\lambda$, and will simplify considerably in the large--$\lambda$ limit. The present analysis is needed to encompass the case of the heterotic $SO(16) \times SO(16)$ model of~\cite{so1616}, for which $\lambda=\frac{5}{3}$, but also other scenarios with $\lambda<1$ that manifest themselves, for example, in the presence of Scherk--Schwarz compactifications~\cite{ss}, where $\lambda=\frac{2}{3}$, as reviewed in~\cite{ssb_review}. As we shall see, the $e$-fold ranges compatible with Planck data~\cite{Planck, forecasts} move toward lower values of $N$ for milder potentials and toward larger values for steeper ones.   
The final section is devoted to three--point correlators involving gravitons, which are adapted to the present scenarios extending the original work in~\cite{maldacena}. They acquire oscillatory features that originate from the underlying bounce dynamics, but lack the additional turning--point contributions, as was anticipated in~\cite{ms}.

%%%%%%%%%%%%%%%%%%%%%%%%%%%%%%%%%%
\section[Climbing phenomenon and scalar three-point functions]{\sc  Climbing Phenomenon and Scalar 3-point Functions}\label{sec:summary}
%%%%%%%%%%%%%%%%%%%%%%%%%%%%%%%%%%%%%
The action for the models of interest describes a scalar field minimally coupled to gravity, with an exponential potential term \( V \), 
\beq
V \ = \ \frac{T}{2 \kappa^2}\ e^{\sqrt{6}\,\lambda\,\phi} \ , \label{potphi}
\eeq
so that
\beq
    { \cal S} \ = \ \int d^4 x \sqrt{-g}\left[\frac{R}{2\,\kappa^2}\ - \ \frac{1}{2}\,g^{\mu \nu }\partial_{\mu}\phi\,\partial_{\nu}\phi\ -\ V(\phi)\right] \ . \label{eq:2der_act}
\eeq
The focus is on a class of four-dimensional, spatially flat metrics of the form
\beq
    ds^2 \ =\  - \ e^{2B(\xi)}d\xi^2 \ + \ e^{2A(\xi)}d\vec{x} \cdot d \vec{x}\ ,
    \label{eq:metric}
\eeq
where \( \xi \) is a "parametric time", chosen in such a way as to simplify the equations of motion. For potentials that never vanish, one can make the gauge choice~\cite{dm}
\beq
    V(\phi)\,e^{2B}\ = \ \frac{M^2}{2\,\kappa^2} \ , \label{eq:gauge_choice}
\eeq
where \( M \) is a scale that can be conveniently identified with the string tension \( T \). Combining this gauge choice with the redefinitions
\beq
    \tau \ = \ \xi \,M \sqrt{\frac{3}{2}} \ , \qquad  \varphi \ = \  \kappa \,\phi\, \sqrt{\frac{3}{2}} \ , \qquad A \ = \ \frac{1}{3}\,a \ ,
    \label{eq:refes}
\eeq
leads to the Hamiltonian constraint for an expanding Universe
\beq\frac{da}{d\tau} \ =\ \sqrt{1\ +\ \left(\frac{d\varphi}{d\tau}\right)^2} \ , 
\eeq
and to the scalar equation
\beq
    \frac{d^2\varphi}{d\tau^2}\ +\ \frac{d\varphi}{d\tau}\sqrt{1\ +\ \left(\frac{d\varphi}{d\tau}\right)^2}\ +\ \frac{1}{2V}\frac{\partial V}{\partial \varphi}\left[1\ +\ \left(\frac{d\varphi}{d\tau}\right)^2\right] \ = \ 0 \ . \label{eq:eqs_gen}
\eeq
In terms of $\varphi$, the potential~\eqref{potphi} becomes
\beq
    V(\varphi)\ = \ \frac{T}{2 \kappa^2}\ e^{2 \,\lambda\,\varphi}  \ , \label{eq:Vgamma}
\eeq
where \( T \) is the string tension, and in this case eq.~\eqref{eq:eqs_gen} becomes
\beq
    \frac{d^2\varphi}{d \tau^2}\ +\ \frac{d\varphi}{d \tau}\sqrt{1\ + \ \left(\frac{d\varphi}{d \tau}\right)^2}\ + \ \lambda\left[1\ + \ \left(\frac{d\varphi}{d \tau}\right)^2\right] \ = \ 0\ .
    \label{eq:eqmot}
\eeq

The solutions of this equation exhibit the peculiar climbing behavior~\cite{climbing}. Without loss of generality, one can restrict the attention to positive values of $\lambda$, up to a reflection, and then:
\begin{itemize}
    \item for \( 0 < \lambda < 1 \) there are two classes of solutions. In the first, the scalar emerges from the initial singularity while climbing the exponential barrier up to a turning point, before reverting its motion and starting a final descent. These are called \textit{climbing solutions}. In the other class of solutions, called \textit{descending}, the scalar emerges from the initial singularity while descending the exponential barrier. The first class of solutions can completely unfold within a weak--coupling regime; 
    \item for \( \lambda \geq 1 \) the descending solutions disappear, leaving only the climbing ones.
\end{itemize}
Interestingly, the \( 0'B \) and \( Usp(32) \) orientifold models of~\cite{susy95} and~\cite{usp32}  have $\lambda=1$ and the \( SO(16) \times SO(16) \) model of~\cite{so1616} has $\lambda=\frac{5}{3}$, so climbing is the only option in all these cases. This characteristic behavior, common to these models, is referred to as the \textit{climbing phenomenon}.

The solutions with $\lambda<1$ approach the Lucchin--Matarrese  attractor~\cite{lm} at late parametric times \( \tau \), which corresponds to
\beq
    \varphi(\tau)\ =\ \varphi_0 \ - \ \frac{\lambda \,\tau}{\sqrt{1-\lambda^2}}\ , \qquad a(\tau)\ =\ \frac{\tau}{\sqrt{1-\lambda^2}} \ ,
    \label{eq:critsol}
\eeq
and thus to a constant limiting speed in the coordinate system determined by eq.~\eqref{eq:gauge_choice},
The limiting form of the metric in conformal time,
\beq
    ds^2 \ = \  \left[\frac{\sqrt{6(1-\lambda^2)}}{M(1-3\lambda^2)} \ \left(-\eta\right) \ e^{-\lambda \varphi_0}\right]^{\frac{2}{1-3\lambda^2}}\Big(-\ d\eta^2\ + \ d \vec{x} \cdot d\vec{x}\Big)\ , 
    \label{eq:lmmetric}
\eeq
approaches the de Sitter one when \( \lambda \ll 1 \). This limiting behavior disappears precisely when the climbing behavior sets in.

This present work is devoted to characterizing the cosmological perturbations to the three--point functions in these backgrounds, complementing the recent results on scalar perturbations in~\cite{ms}. In that paper, the calculation was carried out using the solutions of the background dynamics for the ``critical'' \( \lambda = 1 \) choice:
\beq 
    \varphi(\tau)\ = \ \varphi_0 \ + \ \frac{1}{2}\left(\log \tau \ - \ \frac{\tau^2}{2}\right) \ , \qquad a(\tau)\ = \ \frac{1}{2}\left(\log \tau \ + \ \frac{\tau^2}{2}\right)\ . \label{eq:critical_beh}
\eeq
We had in mind potentials where a "hard" barrier with \( \lambda = 1 \) arising from the orientifold implementation of SUSY breaking was accompanied by a second "mild" contribution capable of sustaining a slow-roll inflationary phase during the final descent. The potential was then implicitly of the form
\beq
    V \ = \ \frac{M^2}{2 \kappa^2}\ e^{2 \,\lambda\, \varphi} \ + \ v_0 \,e^{2 \,\gamma\,\varphi} \ , 
    \label{eq:2exp}
\eeq
with $\lambda=1$ and $\gamma<1$, low enough to grant an inflationary phase. As explained, for example, in~\cite{dkps}, this second term could originate from lower-dimensional branes, and more generally one could combine the hard exponential with more general inflationary potentials, like the celebrated Starobinsky term. One of the goals of the present work is to show how the results change if one relaxes the hypothesis of using the orientifold value $\lambda=1$ for the hard exponent.
To this end, one must rely on the background solutions of the one--exponential potential for general values of \( \lambda \). In particular, the solutions
\bea
    && e^{\lambda \varphi}=e^{\lambda \hat{\varphi}_0}\left[\frac{\sinh{\frac{\tau}{2}\sqrt{1-\lambda^2}}}{\frac{1}{2}\sqrt{1-\lambda^2}}\right]^{\frac{\lambda}{\lambda+1}}\left[\cosh{\frac{\tau}{2}\sqrt{1-\lambda^2}}\right]^{-\frac{\lambda}{1-\lambda}}\ , \nonumber \\
    && e^{\frac{a}{3}}=\left[\frac{\sinh{\frac{\tau}{2}\sqrt{1-\lambda^2}}}{\frac{1}{2}\sqrt{1-\lambda^2}}\right]^{\frac{1}{3(\lambda+1)}}\left[\cosh{\frac{\tau}{2}\sqrt{1-\lambda^2}}\right]^{\frac{\lambda}{3(1-\lambda)}} \ .
    \label{eq:eqs_lno1}
\eea
apply to the range \( 0 < \lambda < 1 \), while 
\bea
    && e^{\lambda \varphi}=e^{\lambda \hat{\varphi}_0}\left[\frac{\sin{\frac{\tau}{2}\sqrt{\lambda^2-1}}}{\frac{1}{2}\sqrt{\lambda^2-1}}\right]^{\frac{\lambda}{\lambda+1}}\left[\cos{\frac{\tau}{2}\sqrt{\lambda^2-1}}\right]^{\frac{\lambda}{\lambda-1}}\ , \nonumber \\
    && e^{\frac{a}{3}}=\left[\frac{\sin{\frac{\tau}{2}\sqrt{\lambda^2-1}}}{\frac{1}{2}\sqrt{\lambda^2-1}}\right]^{\frac{1}{3(\lambda+1)}}\left[\cos{\frac{\tau}{2}\sqrt{\lambda^2-1}}\right]^{-\frac{\lambda}{3(\lambda-1)}} 
    \label{eq:eqs_lno2}
\eea
apply to the complementary range $\lambda>1$. 
The additional constant terms compared to the expressions in~\cite{climbing} grant, in all cases, a smooth limiting behavior as $\lambda\to 1$. As we shall see, the non--analytic behavior of these expressions will disappear in the resulting amplitudes: the matching condition used to link the dynamical Hubble parameter at the turning point to its inflationary value $\mathcal{H}$, as was done for the $\lambda = 1$ amplitudes in ~\cite{ms}, eliminates all problematic terms.

The main cosmological lesson to be drawn from these models was described in~\cite{dkps}. The pre--inflationary climbing phase lowers the corresponding Mukhanov-Sasaki potential compared to the standard quasi-de Sitter shape, with a consequent depression of the low--$\ell$ end of the CMB power spectrum in the presence of short--enough inflationary epochs granting us access to the early deceleration. One can account for this new feature including a constant \( \Delta \) in the Mukhanov--Sasaki equation,
\beq
    v''_k(\eta)\ +\ \left(k^2\ + \ \Delta^2 \ - \ \frac{\nu^2\ - \ \frac{1}{4}}{\eta^2}\right)v_k(\eta)\ = \ 0 \ ,
    \label{eq:mseq}
\eeq
and the deformed power spectrum then takes the form
\beq
    P(k) \ = \ A \ \frac{\left(\frac{k}{k^\star}\right)^3}{\left[ \left(\frac{k}{k^\star}\right)^2 \ + \ \left(\frac{\Delta}{k^\star}\right)^2 \right]^{2 - \frac{n_s}{2}}} \label{eq:power_delta} \ ,
\eeq
where the pivot scale \( k^\star \) can be identified with $\Delta$.

Some evidence for the scale \( \Delta \) was found in~\cite{gkmns} in CMB data. The detection level improves up to the 3$\sigma$ level as the mask around the Galactic plane is enlarged, leaving out the region around the Galactic plane, and the sharpest result,
\beq
    \Delta \ = \ \left (0.351 \pm 0.114\right) \times 10^{- 3}\  \mathrm{Mpc}^{- 1} \ , \label{eq:delta_value}
\eeq
is obtained with an open sky fraction \( f_{sky} \simeq 39\% \). The central value of $\Delta$ identifies a distance scale comparable to the Cosmic Horizon, and by retracing the past history of the Universe it identifies, at the onset of inflation, an energy scale~\cite{gkmns}
\beq
    \Delta_\mathrm{inf} \ \simeq \ 3 \times 10^{14} \ e^{N-60} \ \sqrt{\frac{H_\mathrm{inf}}{\mu_{\mathrm{Pl}}}} \ \mathrm{GeV} \ \simeq \ 2 \times 10^{12} \ \ e^{N-60} \ \mathrm{GeV} \ , \label{eq:DeltaH}
\eeq
where \( H_\mathrm{inf} \simeq 10^{14} \ \mathrm{GeV} \), \( \mu_{\mathrm{Pl}} \simeq 2.4 \times 10^{18} \ \mathrm{GeV} \) is the reduced Planck energy, and \( N \) denotes the number of inflationary \( e \)-folds.

\begin{figure}[ht]
     \centering
     \includegraphics[width=95mm]{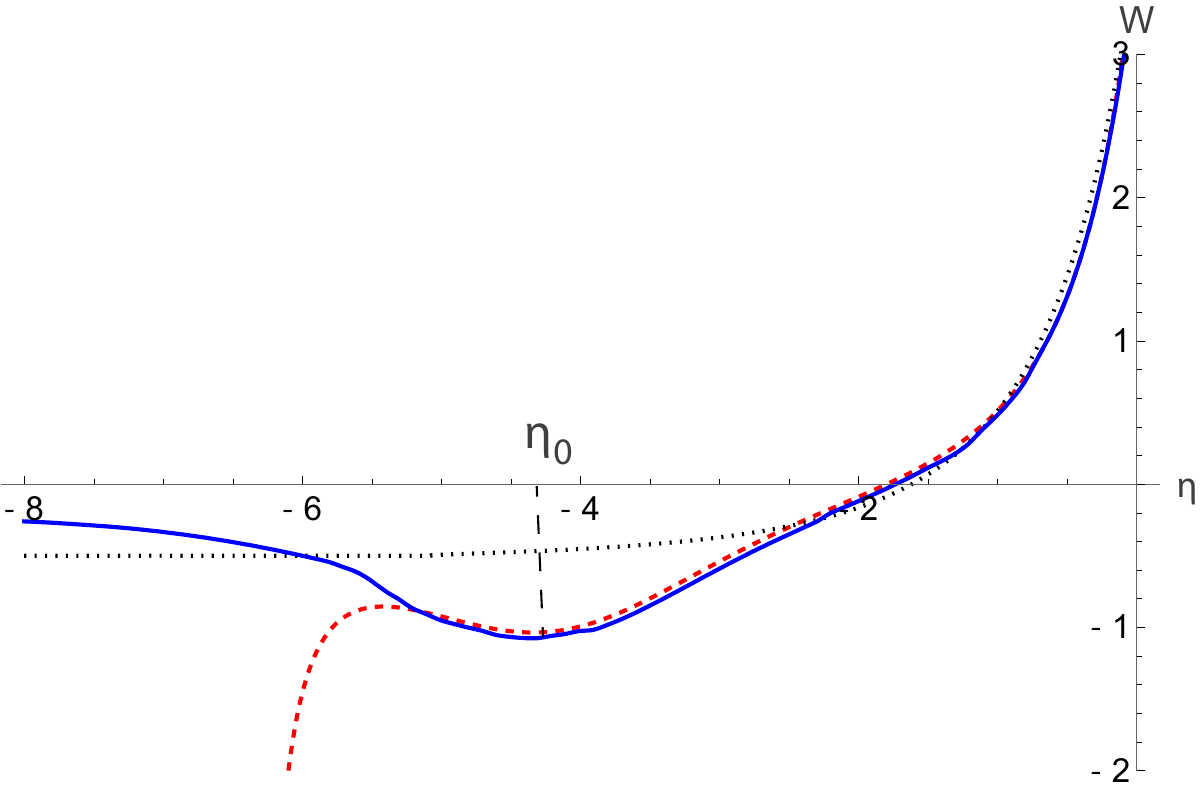}
          \caption{\small The deformed Mukhanov-Sasaki potential that resolves the initial singularity considered in~\cite{ms} (blue, solid), where the Bunch--Davies condition is imposed at $\eta_0$, together with a typical $W$ for a climbing cosmology (red,dashed) and the attractor potential lowered by $\Delta^2$ (black, dotted).}
\label{fig:0}
\end{figure}

The three-point function of curvature perturbations in a climbing-scalar background was investigated in~\cite{ms} for $\lambda=1$, the special value that corresponds to the orientifold models, under the key assumption that the higher-curvature corrections to the string effective action resolve the initial singularity turning it into a bounce. This assumption drew some motivation from the apparent growth of the Mukhanov--Sasaki potential for the climbing scalar as the conformal time $\eta$ decreases, before the tendency is overwhelmed by the singular behavior present in the low--energy dynamics, and has the virtue of eliminating all singular contributions from the three--point amplitude (see Fig.~\ref{fig:0}).

In this context, the Bunch–Davies condition was imposed at the point of maximal contraction $\eta_0$, and the scalar field was in a slow--roll regime throughout its subsequent evolution. Two special values of the conformal time play a special role in the analysis: the first is the value \( \eta_0 \), which corresponds to the bounce, and the second is the value \( \eta_t \), which corresponds to the turning point where the slow--roll parameter $\epsilon$ vanishes. Both $\eta_0$ and $\eta_t$ are given by dimensionless numbers of order one in units of $\Delta^{-1}$. The bispectrum was computed in~\cite{ms} using the Schwinger--Keldysh formalism, and comprises a standard slow--roll contribution, albeit depending on the scale $\eta_0$, which introduces an oscillatory behavior, together with a novel contribution arising from the turning point. This peculiar contribution draws its origin from the vanishing scalar velocity, and can be extracted in an analytically controlled way from residues, thanks to Schwinger--Keldysh deformation of the integration contour, as in ~\cite{ms}, with some minor technical complications, in the present case, due to the more complicated underlying dynamics for $\lambda \neq 1$ of eqs.~\eqref{eq:eqs_lno1} and \eqref{eq:eqs_lno2}. 

For \( \eta \geq  \eta_0 \), the mode functions for scalar perturbations are of the slow-roll form:
\beq
    v_\omega(\eta)\ =  \ - \ \frac{\sqrt{\pi}}{2}\sqrt{-\eta}\ H_{\frac{3}{2}}^{(1)}(-\omega \eta) \ = \  \frac{1}{\sqrt{2 \,\omega}} \left(1 \ - \ \frac{i}{\omega\,\eta} \right)\ e^{-i\,\omega\,\eta} \ , \label{eq:vlargenegeta}
\eeq
but with the frequency deformed by $\Delta$, so that
\beq
    \omega = \sqrt{k^2+\Delta^2} \ .
    \label{eq:omega}
\eeq
This leads to the Green functions
\beq
    G_{\zeta}(\eta,\omega) \ = \ \frac{{\cal H}^2}{4 \,M_p^2 \, \sqrt{3 \, \gamma^2 \, \epsilon (\eta)} \ \omega^{3}} \ e^{i \, \omega\,\eta} \left(1\ - \ i\, \omega\,\eta \right)  \ ,
    \label{eq:green}
\eeq
where $\epsilon$, the slow-roll parameter, is determined by the microscopic dynamics around the turning point according to
\beq
    \epsilon \ = \ \frac{3\,\left(\frac{d\varphi}{d \tau}\right)^2}{1 \ +\  \left(\frac{d\varphi}{d \tau}\right)^2} \ ,
    \label{eq:slowroll}
\eeq
and $3\gamma^2$, the slow-roll value, is  readily recovered away from the turning point.
Under these assumptions, the bispectrum was computed in~\cite{ms} for $\lambda=1$, using the standard cubic action for curvature perturbations in the uniform density gauge introduced by Maldacena. This choice is not merely conventional: the map between the uniform density gauge and the \( \varphi \)-gauge becomes singular at the turning point, making the latter ill-defined in the presence of a turning point. Measurements deal, after all, with curvature perturbations, which justifies in our opinion the choice of focusing on them.

The three-point function results from contributions associated with the different interaction terms in the cubic action obtained after the redefinition in~\cite{maldacena}, after eliminating terms proportional to the equations of motion. The ``principal part'' of the bispectrum, evaluated in the equilateral configuration that maximizes the contributions, is along the lines of the standard slow-roll result, but for an important modification, the presence of mild, \( \Delta \)-dependent, oscillations. It reduces to Maldacena's result if the oscillatory contributions associated with the initial time \( \eta_0 \) are left out, and no sizable signal enhancement is present in these terms.

The most interesting contribution to the bispectrum arises from the turning point, where the slow-roll parameter \(\epsilon\) vanishes. This regime was analyzed in~\cite{ms} for $\lambda=1$, by exploiting the deformed integration contour of the Schwinger--Keldysh formalism, which determines a finite contribution even at this point of strong coupling via a corresponding residue. The ``hard'' exponential term dominates near the turning point, and the result is thus determined by the hard-exponential background solution in parametric time. The resulting three-point function exhibits oscillatory features similar to those present in the principal--part slow-roll contribution, which are also controlled by the scale \(\Delta\). However, the background dynamics introduces in the result different powers of the inflationary Hubble parameter, which translate into a non-trivial dependence on the number of e-folds \(N\). The non-linearity parameter \(f_{NL}\) thus reveals a typical enhancement of about two orders of magnitude within a narrow $e$--fold window \(62 < N < 66\), which could potentially lead to an observable signal still lying within current observational bounds. Outside this range the signal becomes too large: this was a first instance, to the best of our knowledge, where the analysis of inflationary perturbations pointed to such a narrow $e$-fold range.

In the following sections, we shall complement the analysis in two directions. After a brief review of previous results, we shall first describe the three--point amplitude for curvature perturbations for different values of $\lambda$. We shall then turn to amplitudes involving gravity modes. As we anticipated in~\cite{ms}, these do not include turning--point contributions and corresponding enhancements, but their explicit forms represent nevertheless a useful completion of the analysis.

%%%%%%%%%%%%%%%%%%%%%%%%%%%%%%%%%%
\section[Scalar 3-point functions with]{\sc Scalar 3-point functions with \texorpdfstring{$\lambda \neq 1$} \ \ }\label{sec:lambdanotone}
%%%%%%%%%%%%%%%%%%%%%%%%%%%%%%%%%%%%%

The extension of the previous results to the range $\lambda>1$ was not considered in~\cite{ms}, but is needed to cover the case of the $SO(16)\times SO(16)$ heterotic string~\cite{so1616}, and the behavior for $\lambda<1$ can also serve to address other settings, including Scherk--Schwarz compactifications~\cite{ss}, where the leading exponential potential is subcritical and yet not flat enough to sustain inflation by itself, as explained in Section 17 of~\cite{ssb_review}.

The analysis for $\lambda \neq 1$ follows the same steps as in~\cite{ms}, but is somewhat more involved since it rests on~eqs.~\eqref{eq:eqs_lno1} and~\eqref{eq:eqs_lno2}, rather than on the simpler critical background of eqs.~\eqref{eq:critical_beh} that applies to the orientifold models of~\cite{susy95} and~\cite{usp32}. As in that case, we shall be able to absorb a constant dilaton shift by matching the dynamical Hubble parameter at the turning point and the corresponding slow--roll inflationary value $\mathcal{H}$. As in~\cite{ms}, we shall also assume that the inflationary value is quickly recovered away from the turning point, and we shall restrict our attention again to the equilateral configuration, which yields the largest signal in all cases. 

The new non--vanishing contributions from the turning point for $\lambda \neq 1$ are
\bea
    \langle O_1 \rangle_{\text{t}}^{\lambda} &=& -\frac{3 \, \pi \,  \mathcal{H}^5 \, 2^{\frac{\lambda}{3 \left(\lambda ^2-1\right)}} \left(\lambda^{-1}+1\right)^{\frac{1}{6(1- \lambda) }} \, \lambda^{\frac{6 \lambda + 5}{6(\lambda + 1)}}  \left(\lambda +1\right)^{-\frac{1}{6 (\lambda +1)}}}{16 \, \gamma ^3 \, \eta_t^2 \, {M_p}^4 \, \omega ^9} \nonumber\\ &\times& \left[\eta_t \, \omega  \left(\eta_t^2 \, \omega ^2-3\right) \sin (3 \, \eta_t \, \omega )+\left(3 \, \eta_t^2 \, \omega ^2-1\right) \cos (3 \, \eta_t \, \omega )\right] \ , \nonumber\\
    \langle O_3 \rangle_{\text{t}}^{\lambda} &=& -\frac{3 \, \pi \,  \mathcal{H}^5 \, 2^{\frac{\lambda}{3 \left(\lambda ^2-1\right)}} \left(\lambda^{-1}+1\right)^{\frac{1}{6(1- \lambda) }} \, \lambda^{\frac{6 \lambda + 5}{6(\lambda + 1)}}  \left(\lambda +1\right)^{-\frac{1}{6 (\lambda +1)}}}{16 \, \gamma ^3 \, \eta_t^2 \, {M_p}^4 \, \omega ^9} \label{eq:turn_contributions2} \\ &\times& \left[\eta_t \, \omega  \left(\eta_t^2 \, \omega ^2-3\right) \sin (3 \, \eta_t \, \omega )+\left(3 \, \eta_t^2 \, \omega ^2-1\right) \cos (3 \, \eta_t \, \omega )\right] \ ,\\
    \langle O_4 \rangle_{\text{t}}^{\lambda} &=&  -\frac{\pi \, \mathcal{H} }{288 \, \gamma ^3 \, \eta_t^6 \, \lambda^3 \, {M_p}^4 \, \omega ^9} \nonumber \\ &\times& \bigg\{ 90 \, \eta_t^3 \, \mathcal{H}^3 \, \lambda ^4 \left[ \eta_t \, \omega \left(\eta_t^2 \, \omega^2-3\right) \sin (3 \, \eta_t \, \omega ) + \left(3 \, \eta_t^2 \, \omega ^2-1\right) \cos (3 \, \eta_t \, \omega ) \right] \nonumber \\  &+& 2^{\frac{\lambda }{1- \lambda ^2}} \lambda^{\frac{2 \lambda^2-3}{\lambda ^2-1}} (\lambda +1)^{\frac{\lambda }{\lambda ^2-1}} \Big[ 2^{\frac{3-2\lambda - 3 \lambda^2}{3(1-\lambda^2)}} \, \lambda ^{-\frac{2}{3(1- \lambda ^2)}} \, (\lambda +1)^{-\frac{2\lambda }{3 \left(\lambda ^2-1\right)}} \, \eta_t^2 \, \mathcal{H}^2 \nonumber \\ &\times& \big[(3 \, \eta_t^4 \, \omega^4 \, + \, 9 \, \lambda ^2 \left(3 \, \eta_t^2 \, \omega ^2 \,- \, 1\right) -15 \, \eta_t^2 \, \omega ^2+4) \cos (3 \, \eta_t \, \omega) \nonumber \\ &+& \eta_t \, \omega  \left(9 \, \lambda ^2 \left(\eta_t^2 \, \omega ^2 \, - \, 3\right)-10 \, \eta_t^2 \, \omega^2 \, + \, 12\right) \, \sin (3 \, \eta_t \, \omega )\big]\nonumber\\ &-& 2^{-\frac{\lambda}{3(1- \lambda ^2)}} \lambda ^{-\frac{1}{3(1-\lambda^2)}} (\lambda +1)^{-\frac{ \lambda }{3 \left(\lambda ^2-1\right)}} \, 3 \, \eta_t \, \mathcal{H} \nonumber \\ &\times& \big[ 3 \, \eta_t \, \omega \left(\eta_t^4 \, \omega ^4 \, - \, 5 \, \eta_t^2 \, \omega ^2 \,+ \, 4\right) \sin (3 \, \eta_t \, \omega ) \, + \left(9 \, \eta_t^4 \, \omega ^4 \, - \, 17 \, \eta_t^2 \omega ^2+4\right) \cos (3 \, \eta_t \, \omega ) \big]\nonumber\\ &-& \, \big[\eta_t \, \omega\left(9 \, \eta_t^4 \, \omega ^4 \,- \,65 \, \eta_t^2 \, \omega ^2\, + \, 60\right) \sin (3 \, \eta_t \, \omega )\, +\left(33 \, \eta_t^4 \,\omega ^4 \,- \,81 \, \eta_t^2 \, \omega ^2 \,+ \, 20\right) \cos (3 \, \eta_t \, \omega )\big] \Big] \bigg\}\ ,  \nonumber
    \label{eq:newamp}
\eea
while
\bea
\langle O_2 \rangle_{\text{t}}^{\lambda} &=& 0 \ , \nonumber \\
\langle O_5 \rangle_{\text{t}}^{\lambda} &=& 0 \ ,
\eea
as was the case for $\lambda=1$.

Note that each term has a smooth limit as $\lambda \to 1$, and in fact the preceding expressions apply to the two intervals $\lambda > 1$ and $0 < \lambda < 1$, as we had anticipated.
The results for $\lambda=1$ presented in~\cite{ms} can be recovered noting that
\beq
    \lim_{\lambda \to 1}2^{\frac{\lambda}{3 \left(\lambda ^2-1\right)}} \, \left(\lambda^{-1}+1\right)^{\frac{1}{6(1- \lambda) }}=(2 \, e)^{\frac{1}{12}} \ ,
    \label{eq:limit}
\eeq
so that, for example, 
\beq
    \langle O_1 \rangle_{t}^{\lambda} \to -\ \frac{3\,\pi\, \mathcal{H}^5 \,{e}^\frac{1}{12}}{16\, \gamma^3 \,\eta_t^2\, {M_p}^4 \omega^9} \  \Big[\eta_t \omega  \left(\eta_t^2 \omega ^2-3\right) \sin (3 \eta_t \omega )\ + \ \left(3 \eta_t^2 \omega ^2-1\right) \cos (3 \eta_t \omega )\Big]  \ , 
    \label{eq:limitingO1}
\eeq 
and similar considerations apply to the fourth, more complicated, amplitude.

The fourth amplitude is indeed more complicated, since its contributions arise from a fourth-order pole, but the $\lambda$--dependence of the (identical) first and third contributions is simply an overall factor, so that
\beq
    \langle O_{1,3} \rangle_{\text{t}}^{\lambda} \, = \, h(\lambda) \, \langle O_{1,3} \rangle_{\text{t}} \ ,
    \label{eq:specrel}
\eeq
with
\beq
    h(\lambda) \, = \, e^{- \frac{1}{12}} \,  2^{\frac{\lambda }{3 \left(\lambda ^2-1\right)}} \, \left(\lambda^{-1}+1\right)^{\frac{1}{6(1-\lambda) }} \, \lambda^{\frac{6 \lambda +5}{6(\lambda + 1)}} \,  (\lambda +1)^{-\frac{1}{6 (\lambda +1)}} \ .
    \label{eq:specrel2}
\eeq
For \( \lambda \) slightly below one and beyond, \( h(\lambda) \) grows linearly with \( \lambda \). For \( 0 < \lambda < 1 \), the first and third contributions are thus reduced with respect to the critical solution, while for models with \( \lambda > 1 \), and thus for the $SO(16) \times SO(16)$ model of~\cite{so1616}, they are enhanced.

\begin{figure}[ht]
    \centering
    \includegraphics[width=70mm]{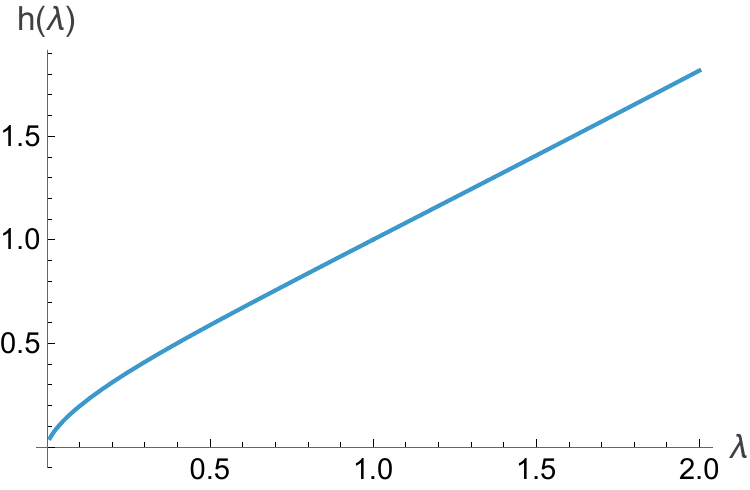}
    \caption{\small $h(\lambda)$ around the critical value $\lambda_c=1$ and beyond.}
    \label{fig:2}
\end{figure}

Interestingly, even the more complicated \( O_4 \) contribution exhibits a simple, if approximate, scaling behavior in \( \lambda \), within the region of interest, where $\lambda$ is below about 1.7, so that the full amplitude is well described by
\beq
    \mathcal{A}_3^{\lambda}\simeq -\frac{\pi \,  \mathcal{H}^3 \, \lambda \,  (2 \eta_t \, \mathcal{H} \, + \, 1) (3 \, \eta_t \, \mathcal{H} \, + \, 1) \left(\eta_t \, \omega  \left(\eta_t^2 \, \omega^2 \,- \, 3\right) \sin (3 \, \eta_t \, \omega) \, + \, \left(3 \, \eta_t^2 \, \omega^2 \, - \, 1\right) \cos(3 \, \eta_t \, \omega)\right)}{16 \, \gamma^3 \, \eta_t^4 \, M_{p}^4 \, \omega ^9} \ .
    \label{eq:biglamb}
\eeq

\begin{figure}[ht]
    \centering
    \begin{tabular}{ccc}
    %\mbox{graphic} & \mbox{table} \\
    \includegraphics[width=70mm]{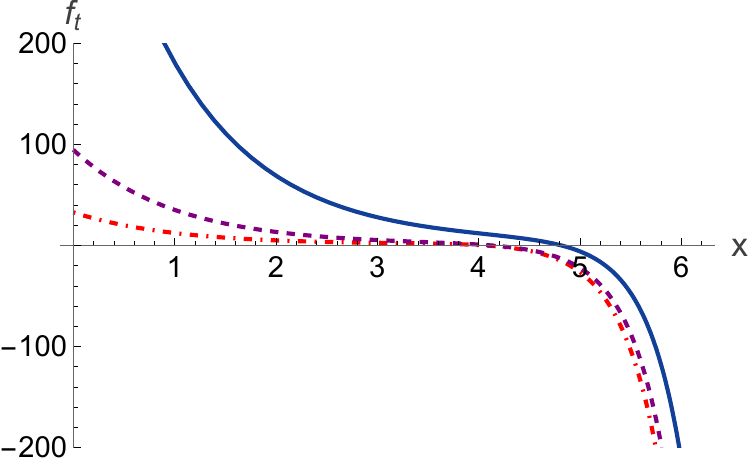} \quad  &
    \includegraphics[width=70mm]{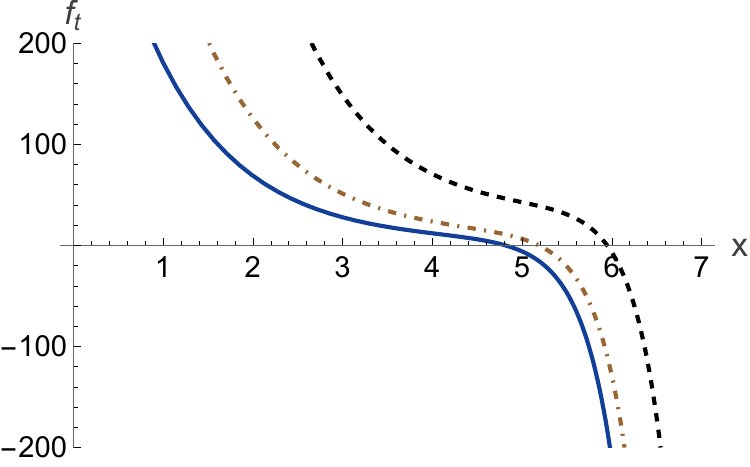}  \\
    \end{tabular}
    \caption{\small Left panel: the turning--point contribution $f_t$ to $f_{NL}$ as a function of the number of e-folds $x=N-60$ for $\lambda = 0.02$ (red, dot-dashed), for $\lambda = 0.1$ (purple, dashed) and $\lambda = \lambda_c = 1$ (dark blue, solid), for $k=1.2 \, \Delta$. Right panel: the turning--point contribution $f_t$ to $f_{NL}$ as a function of the number of e-folds $x=N-60$ for $\lambda = \lambda_c = 1$ (dark blue, solid), for $\lambda = 2$ (brown, dot-dashed) and $\lambda = 6$ (black, dashed) for $k=1.2 \, \Delta$.  The results displayed correspond to the choice $\epsilon=0.03$, with the reference value $\eta_t = -\frac{5}{\Delta}$.}
    \label{fig:ftlamb}
\end{figure}

\begin{figure}[ht]
    \centering
    \includegraphics[width=70mm]{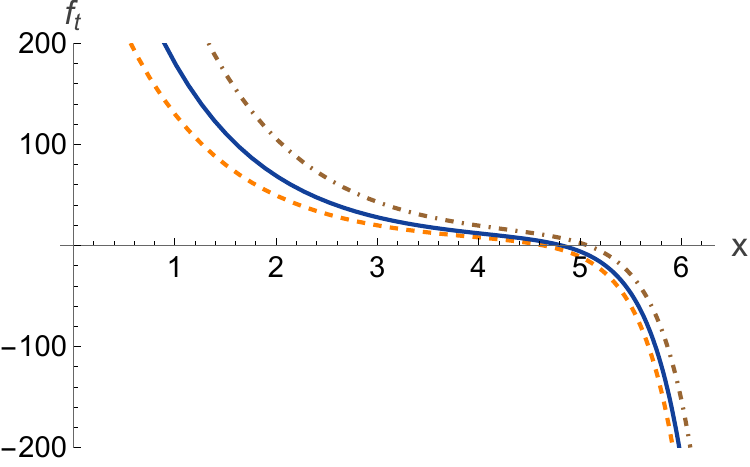}
    \caption{\small The turning--point contribution $f_t$ to $f_{NL}$  as a function of the number of e-folds $x=N-60$ for $\lambda = 2/3$ (orange, dashed), for $\lambda = \lambda_c = 1$ (dark blue, solid) and $\lambda = 5/3$ (brown, dot-dashed), for $k=1.2 \, \Delta$.  The results displayed correspond to the choice $\epsilon=0.03$, with the reference value $\eta_t = -\frac{5}{\Delta}$.}
    \label{fig:ftlamb2}
\end{figure}

Turning to the non--linearity parameter
\beq
    f_{NL}(k)\ = \ 40 \, \frac{\gamma^4 \, M_{p}^4\,\Delta^{6}}{ \, \mathcal{H}^4}  \left[\left(\frac{k}{\Delta}\right)^2 + 1\right]^{2 \, \nu} \, \mathcal{A}_3^{\lambda} \ ,
    \label{eq:fnl}
\eeq
we can now illustrate in Fig.~\ref{fig:ftlamb} how the result depends on the number of inflationary $e$-folds. To this end, we focus on modes with $k = 1.2 \, \Delta$ and 
$\eta_t = -\frac{5}{\Delta}$, as was done in~\cite{ms}. 
Other modes with larger values, such as $k = 2 \, \Delta$, exhibit a similar behavior (in absolute value).
Although we do not consider values of $\lambda$ that are too small, since this would contradict the assumption that the effect of the tadpole exponential only dominates around the turning point, 
the left panel indicates that Scherk--Schwarz models, which contain exponential terms of this type with $\lambda=\frac{2}{3}$, select a lower interval for the number of $e$-folds, and the region where $f_{NL}$ actually remains below about $100$ becomes larger as $\lambda$ is reduced.
On the other hand, the right panel indicates that larger values of $\lambda$ (here chosen beyond the region of interest to enhance the effect that we are trying to highlight) shift the central value of the acceptable window for $x$ toward larger values and thus demand larger numbers of $e$-folds. However, the difference with respect to the critical case is not particularly significant for the value $\lambda = \frac{5}{3}$ of the heterotic scenario, as can be clearly seen in Fig.~\ref{fig:ftlamb2}.
Altogether, there are small, opposite displacements of the relevant window for the acceptable number of $e$-folds in the heterotic and Scherk--Schwarz cases, with respect to what happens for the orientifolds.

\begin{figure}[ht]
    \centering
    \begin{tabular}{ccc}
    %\mbox{graphic} & \mbox{table} \\
    \includegraphics[width=50mm]{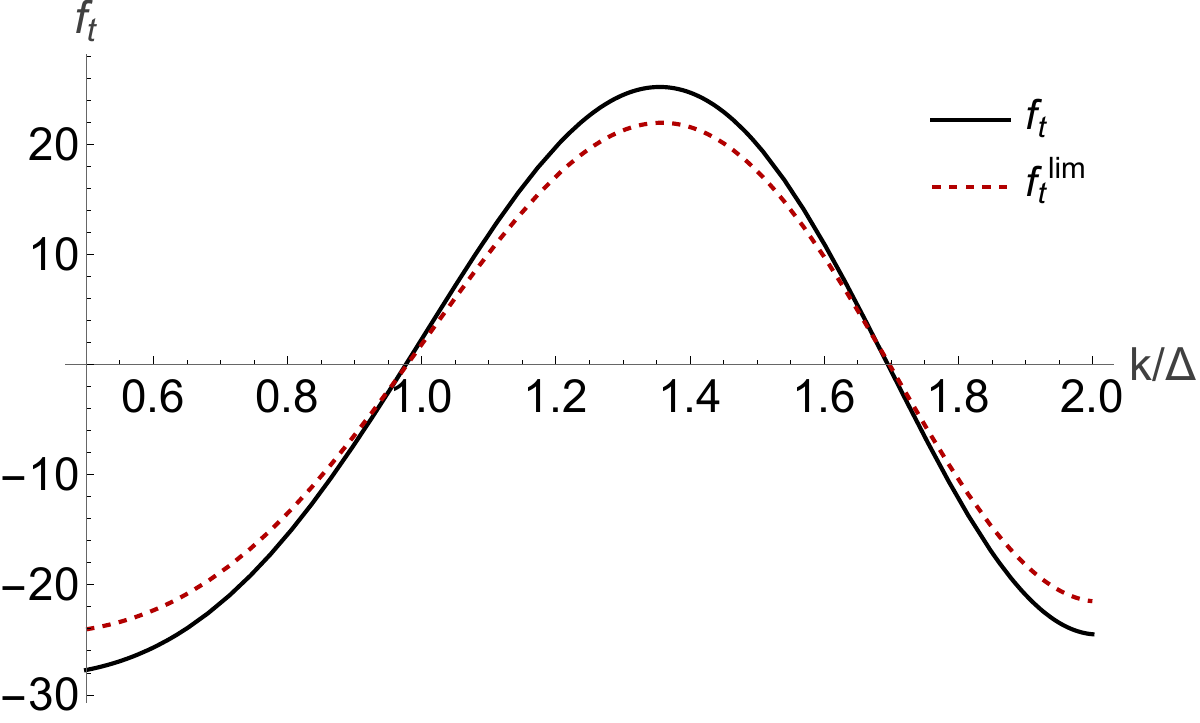} \quad  &
    \includegraphics[width=50mm]{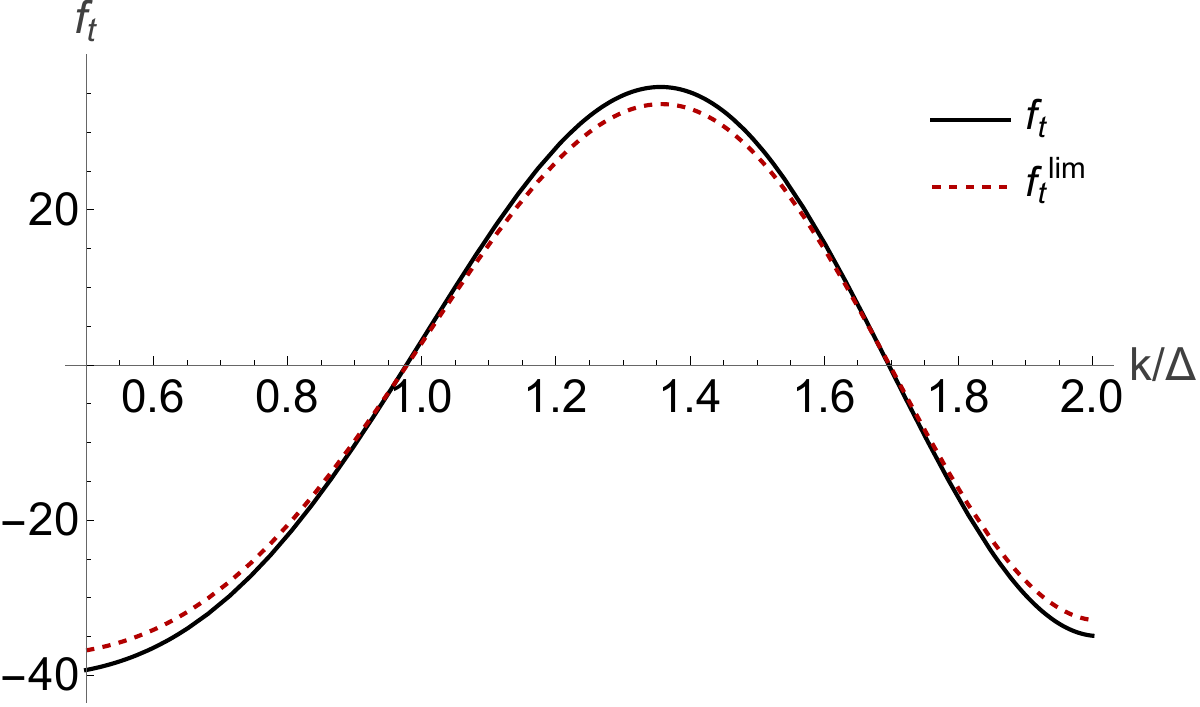} \quad  &
    \includegraphics[width=50mm]{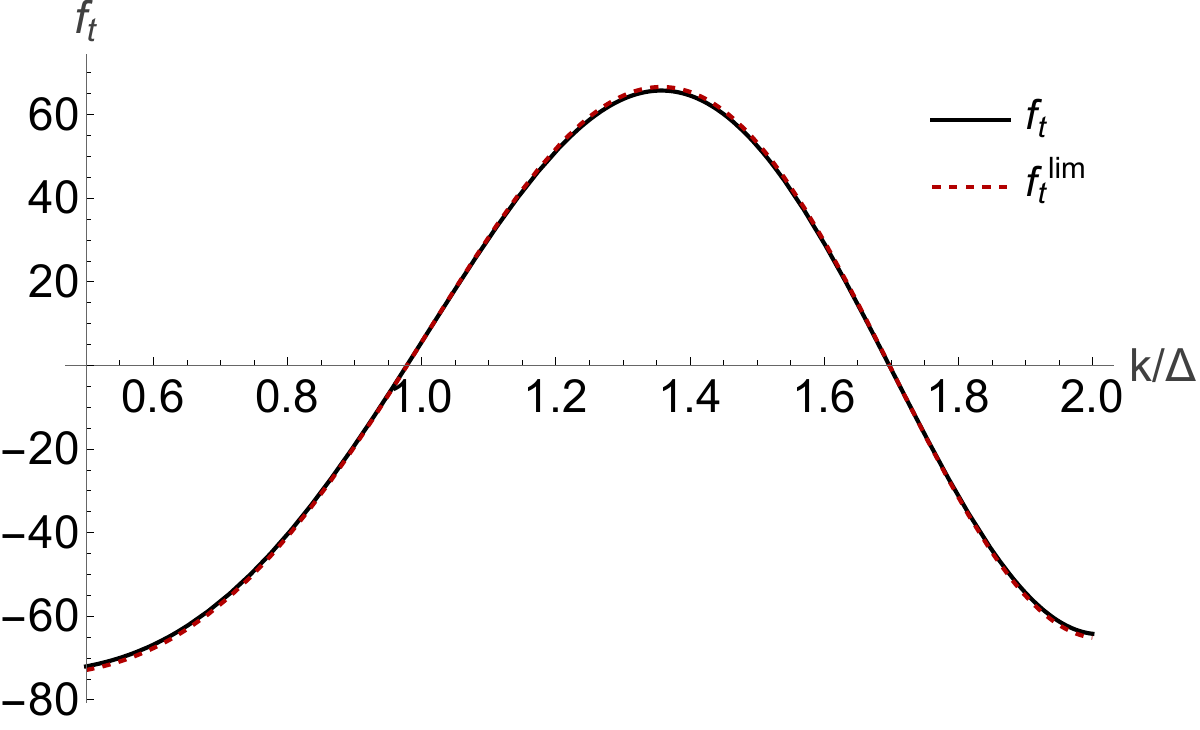}  \\
    \end{tabular}
    \caption{\small $f_{t}$ (black solid) vs $f_{t}^{\text{lim}}$ (red, dashed) as functions of $\frac{k}{\Delta}$. Left panel: $\lambda = 2/3$; central panel: $\lambda = \lambda_c = 1$; right panel: $\lambda = 1.65$. The results correspond to the choice $\epsilon = 0.03$, with reference values $\eta_t = -\frac{5}{\Delta}$ and $N = 63$.}
    \label{fig:ftformtest}
\end{figure}

We can also test the asymptotic formula~\eqref{eq:biglamb} by comparing the exact values of $f_t$ with the corresponding limiting form $f_t^{\text{lim}}$. Using $N = 63$ as a reference value for the number of e-folds---in the middle of the usual range of interest---we compare in Fig.~\ref{fig:ftformtest} the curves corresponding to three cases: Scherk-Schwarz, orientifolds, and $SO(16) \times SO(16)$ heterotic.
For the Scherk--Schwarz value of $\lambda$ the asymptotic form is already fairly accurate, although it lacks about 15\% of the overall amplitude. In the other cases, the two curves are almost indistinguishable.
%%%%%%%%%%%%%%%%%%%%%%%%%%%%%%%%%%
\section[Tensor 3-point functions]{\sc Tensor 3-point functions}\label{sec:tensor}
%%%%%%%%%%%%%%%%%%%%%%%%%%%%%%%%%%%%%
 
The detailed form of the three-point amplitudes involving graviton modes was not discussed in~\cite{ms}. The purpose of this section is to fill this gap, showing in detail that these correlation functions do not receive contributions from the turning point and displaying the expressions that emerge in the bounce cosmology suggested by the (smoothed) climbing scenario.

Our new ingredient is the Mukhanov–Sasaki equation for the graviton mode functions. In the uniform-density gauge, the quadratic action for tensor perturbations $\gamma_{ij}$ is
\beq
    \mathcal{S}^{(2)}_{\gamma} = \frac{M_p^2}{8} \int \! d^4x \, e^{a} \, \left[\dot{\gamma}_{ij} \, \dot{\gamma}_{ij}
    - \, e^{-2  a/3} \, \partial_l \, \gamma_{ij} \, \partial_l \, \gamma_{ij} \,\right] \ ,
    \label{eq:freegrav}
\eeq
and tensor perturbation can be expanded in plane waves with definite polarization tensors,
\beq
    \gamma_{ij}(\mathbf{x}, t)
    =  \int \frac{d^3 k}{(2\pi)^3}
    \sum_{s = \pm}
    \epsilon^{s}_{ij}(\mathbf{k}) \,
    \gamma^{s}_{k}(t) \,
    e^{i \mathbf{k} \cdot \mathbf{x}} \ ,
    \label{eq:fuorgamma}
\eeq
where
\beq
\epsilon^{s}_{ii}(\mathbf{k}) = 0 \ , \qquad k_i \, \epsilon^{s}_{ij}(\mathbf{k}) = 0  \ , \qquad 
\epsilon^{s}_{ij}(\mathbf{k}) \, \epsilon^{s'}_{ij}(\mathbf{k}) = 2\,\delta_{ss'} \ .
\eeq
For each polarization, one is thus again led to a massless scalar equation. However, there is a crucial difference with respect to the scalar case, since now the Mukhanov–Sasaki variable does not involve any $\epsilon$-dependent factors. For each polarization mode, one can define
\beq
\gamma_{k}=e^{-a/3}u_{k} \ ,
\eeq
and then
\beq
    u_{k}'' \, + \, \left[k^2-\frac{a''}{3} \, - \, \left(\frac{a'}{3}\right)^2\right] \, u_{k} \, = \, 0 \ .
    \label{eq:gammarel}
\eeq

Note that the effective potential does not depend on $\Delta$, in contrast with what happens for curvature perturbations. This feature reflects the assumptions previously made on the evolution of the scale factor from the point of maximal contraction into a standard de Sitter phase, and is justified by the detailed results presented in~\cite{dkps}. The absence of deformations in the Mukhanov-Sasaki tensor equation is consistent with the contextual absence of $\epsilon$-dependent terms in effective action. 
The Mukhanov-Sasaki tensor potential for the climbing backgrounds analyzed in~\cite{dkps} intersects the $\eta$-axis, signaling a deviation from the standard de Sitter form. However, the intersection occurs for values of the conformal time much lower than in the scalar case, and so it is natural to assume that it precedes the bounce when string corrections are taken into account, which makes it irrelevant for the subsequent evolution. Here we shall ignore its effect, which could be simply included turning $k$ into an expression similar to eq.~\eqref{eq:omega}. 

Imposing the Bunch–Davies initial condition at the time of maximal contraction and restricting the analysis to the subsequent evolution in conformal time, one can work with the mode functions
\beq
    u_{k}(\eta)\ =  \ - \ \frac{\sqrt{\pi}}{2}\sqrt{-\eta}\ H_{\frac{3}{2}}^{(1)}(-k \eta) \ = \  \frac{1}{\sqrt{2 \,k}} \left(1 \ - \ \frac{i}{k\,\eta} \right)\ e^{-i\,k\,\eta} \ ,
    \label{eq:modef}
\eeq
and the corresponding Green functions
\beq
    G_{\gamma}(0,\eta)=\frac{\mathcal{H}^2}{M_p^2}\frac{1}{2 \, k^3}(1-ik\eta)e^{ik\eta} \ ,
    \label{eq:wightman}
\eeq
so that the resulting two-point function coincides with the standard de Sitter result,
\beq
    \langle\gamma_{\mathbf{k}}^s \, \gamma_{\mathbf{k}'}^{s'}\rangle=(2 \pi)^3 \delta^{(3)}(\mathbf{k}+\mathbf{k}')\frac{1}{2k^3}\frac{\mathcal{H}^2}{M_{pl}^2} \, 2 \,\delta^{s s'} \ .
    \label{eq:twograv}
\eeq

We are now ready to discuss in detail the three-point amplitudes involving graviton modes. Following Maldacena’s original analysis~\cite{maldacena}, we present the results in the following order: two scalars and one graviton, one scalar and two gravitons, and finally three gravitons.
%%%%%%%%%%%%%%%%%%%%%%%%%%%%%%%%%%
\subsection[Two scalars and one graviton]{\sc Two scalars and one graviton}\label{sec:twoscalarsandonegraviton}
%%%%%%%%%%%%%%%%%%%%%%%%%%%%%%%%%%%%%
The relevant portion of the cubic action in the uniform-density gauge is~\cite{maldacena}
\bea
    \mathcal{S}^{(3)}_{\zeta \zeta\gamma} &=& M_p^2 \int d^4x \bigg[
    \epsilon \, e^{a/3} \, \gamma_{ij} \, \partial_i \zeta \, \partial_j \zeta
    + \frac{\epsilon^2}{4} e^{a} \, \partial^2 \gamma_{ij} \,
    \partial_i\!\left(\partial^{-2} \dot{\zeta}\right)
    \partial_j\!\left(\partial^{-2} \dot{\zeta}\right)
    \nonumber\\
    &\quad&
    + \frac{\epsilon^2}{2} e^{a} \, \dot{\gamma}_{ij} \,
    \partial_i \zeta \,
    \partial_j\!\left(\partial^{-2} \dot{\zeta}\right)
    + \hat{f}(\zeta, \gamma)
    \frac{\delta L}{\delta \zeta}\bigg|_1
    + \hat{f}_{ij}(\zeta)
    \frac{\delta L}{\delta \gamma_{ij}}\bigg|_1
    \bigg] \ ,
    \label{eq:cubact1}
\eea
and the terms proportional to the equations of motion can be removed by suitable field redefinitions. These contributions become irrelevant after horizon crossing, as they involve only derivatives that scale with the conformal time and therefore do not affect the present computation, as in~\cite{maldacena}. They will be neglected in what follows.

In analogy with the three-scalar case, one could distinguish principal-part and turning-point contributions. However, a simple $\epsilon$-counting shows that the interactions appearing in~\eqref{eq:cubact1} do not generate any poles, so that there are no turning-point contributions in this case. We can now see this in detail, recalling that each $\zeta$ propagator $G_\zeta$ carries a dynamical factor $\sqrt{\epsilon}$, while the graviton propagator $G_\gamma$ does not involve any $\epsilon$ dependence, as we have seen in eq.~\eqref{eq:twograv}. The leading interaction term contains no time derivatives and therefore scales as
\begin{equation}
    \epsilon \, \frac{1}{\sqrt{\epsilon}} \, \frac{1}{\sqrt{\epsilon}} = \epsilon^0
    \ .
    \label{eq:epslcount}
\end{equation}
where we have taken into account the behavior of eq.~\eqref{eq:green}, which implies indeed the absence of turning-point contributions.

The second interaction term involves two time derivatives acting on $\zeta$. Enclosing within parentheses the pairs of contributions originating from the factors with derivatives and using the symbol $\times$ to build their combinations, one finds 
\begin{equation}
    \epsilon^2 \times \left(\frac{1}{\sqrt{\epsilon}}, \frac{\epsilon '}{\epsilon^{3/2}}\right) \times \left(\frac{1}{\sqrt{\epsilon}}, \frac{\epsilon '}{\epsilon^{3/2}}\right)=\left(\epsilon, \epsilon', \frac{(\epsilon')^2}{\epsilon}\right) \ .
    \label{eq:epslcount2}
\end{equation}
The first two terms clearly do not yield turning-point contributions, but the last term requires a more detailed analysis. However, for the estimate it suffices to restrict the attention to the simpler expression for the critical case $\lambda = \lambda_c = 1$, for which
\begin{equation}
    \frac{(\epsilon')^2}{\epsilon}=\frac{1728 \,  \mathcal{H}^2 e^{\frac{1}{2}-\frac{\tau ^2}{3}} \tau ^{10/3}}{\left(\tau ^2+1\right)^4} \ .
    \label{eq:epslcount3}
\end{equation}
Since there are no poles here at $\tau = 1$, there are again no turning-point contributions.
Finally, the last interaction term contains a single time derivative acting on $\zeta$, leading to the combinations
\begin{equation}
    \epsilon^2 \times \frac{1}{\sqrt{\epsilon}} \times \left(\frac{1}{\sqrt{\epsilon}}, \frac{\epsilon '}{\epsilon^{3/2}}\right)=\left(\epsilon, \epsilon'\right) \ ,
    \label{eq:epslcount4}
\end{equation}
which again do not produce any turning-point contributions.

We therefore conclude that turning-point contributions are completely absent in this sector, as we anticipated in~\cite{ms}. Therefore, the result is entirely determined by the principal part contribution that has an oscillatory behavior depending on the initial time $\eta_0$. Neglecting subdominant terms, the three-point function takes the form
\beq
    \langle \gamma^{s}_{\mathbf{k}_1} \, \zeta_{\mathbf{k}_2} \, \zeta_{\mathbf{k}_3} \rangle
    = (2\pi)^3 \, \delta^{3}\!\left(\sum_i \mathbf{k}_i \right)
    \frac{1}{12 \, \gamma^2 \, k_1^3 \, \omega_2^3 \, \omega_3^3} \,
    \frac{\mathcal{H}^4}{M_{p}^4}\,
    \epsilon^{s}_{ij} \, k_{2}^i \, k_{3}^j \, I_1 \ ,
    \label{eq:amp1}
\eeq
where the integral $I_1$ is defined as
\beq
    I_1 = \mathrm{Re} \left[
    - i \int_{\eta_0}^{0} \frac{d\eta}{\eta^{2}}
    (1 - i k_1 \eta)(1 - i \omega_2 \eta)(1 - i \omega_3 \eta)
    e^{i \vartheta \eta}
    \right] \ ,
    \label{eq:I1u}
\eeq
with $\vartheta \equiv k_1 + \omega_2 + \omega_3$.

In the equilateral configuration, which provides the largest signal, one thus obtains
\bea
    I_1&=&-\frac{\eta_0 \left(k^3+4 k^2 \omega +6 k \omega ^2+6 \omega ^3\right)}{\eta_0 (k+2 \omega )^2} \nonumber \\ 
    &-&\frac{2 \, \eta_0 \, \omega  \left(k^2+3 \, k \, \omega +\omega ^2\right) \cos [\eta_0 (k+2 \, \omega )]+(k+2 \, \omega ) \left(\eta_0^2 \, k \, \omega ^2-k-2 \omega \right) \sin [\eta_0 (k+2 \, \omega )]}{\eta_0 (k+2 \, \omega )^2} . \nonumber
    \label{eq:I1}
\eea
As in the three-scalar case, leaving out the trigonometric terms and taking the limit $\Delta \to 0$, so that $\omega \to k$, only the first line survives, and one recovers Maldacena’s result for this amplitude.
%%%%%%%%%%%%%%%%%%%%%%%%%%%%%%%%%%
\subsection[One scalar and two gravitons]{\sc One scalar and two gravitons}\label{sec:onescalarandtwogravitons}
%%%%%%%%%%%%%%%%%%%%%%%%%%%%%%%%%%%%%
The relevant portion of the cubic action in the uniform-density gauge is~\cite{maldacena}
\bea
    S^{(3)}_{\zeta\gamma\gamma} &=& M_p^2\int d^4x \bigg\{\frac{\epsilon}{8}
    \left[
    e^{a} \, \zeta \, \dot{\gamma}_{ij} \, \dot{\gamma}_{ij}
    + e^{a/3} \, \zeta \, \partial_{l} \, \gamma_{ij} \, \partial_{l} \, \gamma_{ij}
    \right]
    - \frac{\epsilon}{4} \, e^{a} \, \dot{\gamma}_{ij} \, \partial_{l} \, \gamma_{ij} \, \partial_{l} \left(\partial^{-2} \dot \zeta\right)
    \nonumber \\ &-& g_{ij}(\zeta, \gamma) \frac{\delta L}{\delta \gamma_{ij}}\bigg\} \ .
    \label{eq:cubact2}
\eea
All terms proportional to the equations of motion can again be removed by field redefinitions and yield vanishing contributions to the correlator in the super-horizon limit. 

We can now show that no turning-point contributions arise even in this case. For the first two terms, the $\epsilon$-scaling is
\beq
        \epsilon \, \epsilon^{-1/2} = \epsilon^{1/2} \ ,
\eeq
    so that they are manifestly regular. For the third term, the contributions are of the form
\beq
        \epsilon \times \left(\frac{1}{\sqrt{\epsilon}},\frac{\epsilon'}{\epsilon^{3/2}}\right)
        = \left(\sqrt{\epsilon},\frac{\epsilon'}{\epsilon^{1/2}}\right) \ ,
\eeq
    and the potentially dangerous one is the second. However, a closer look reveals that
\beq
        \frac{\epsilon'}{\epsilon^{1/2}}
        = \frac{24 \, \sqrt{3} \, \mathcal{H} \,
        e^{\frac{1}{4}-\frac{\tau ^2}{6}} \, \tau ^{5/3}}
        {(\tau ^2+1)^2} \ ,
\eeq
    with no poles at the turning point.

The absence of turning-point contributions considerably simplifies the analysis, reducing the correlator to the principal-part contribution only. One thus finds
\beq
    \langle \zeta_{\mathbf{k_1}} \, \gamma_{\mathbf{k}_2}^{s_2} \, \gamma_{\mathbf{k}_3}^{s_3}\rangle=(2\pi)^3 \, \delta^{3}\!\left(\sum_i \mathbf{k}_i \right)
    \frac{1}{16\, \omega_1^3 \, k_2^3 \, k_3^3} \,
    \frac{\mathcal{H}^4}{M_{pl}^4}\,
    \epsilon^{s_2}_{ij} \epsilon^{s_3}_{ij} I_2  \ ,
    \label{eq:amp2}
\eeq
and letting
\beq
I_2=I_2^{O_1}+I_2^{O_2}+I_2^{O_3} 
\eeq
and
\beq
\theta = \omega_1 + k_2 + k_3 \ , 
\eeq
the individual contributions read
\bea
    I_2^{O_1}&=& i \, k_2^2 k_3^2 \,\left[\int^0_{\eta_0}d\eta (1-ik_1\eta)e^{i \theta \eta} \,  -\, c.c.\right]
    \nonumber \\
    I_2^{O_2}&=&-i \, \left(\frac{k_1^2-k_2^2-k_3^2}{2}\right)\left[\int^0_{\eta_0} d \eta\frac{(1-i\omega_1 \eta)(1-ik_2 \eta)(1-ik_3 \eta)}{\eta^2}e^{i \theta \eta} \, - \, c.c.\right]
    \nonumber \\  
    I_2^{O_3}&=&-i \left[k_3^2 \, (\mathbf{k_1} \cdot \mathbf{k_2})\int^0_{\eta_0}(1-ik_2 \eta)e^{i \theta \eta} \, - \, c.c.\right]+(k_2 \leftrightarrow k_3) \ .
    \label{eq:I2s}
\eea
In the equilateral configuration, one finally obtains
\bea
    I_2&=&\frac{k^2}{\eta_0 \, (2 \, k+\omega )^2} \Big\{2 \, \eta_0 \, k \left(-5 \, k^2 +k \, \omega +\omega ^2\right) \cos [\eta_0 (2 \, k+\omega )]+\eta_0 \left(18 \, k^3+10 \, k^2 \, \omega +4 \, k \, \omega ^2+\omega ^3\right)\nonumber \\ &-&(2 \, k+\omega ) \left[k \left(\eta_0^2 \, k \left(4 \, k-\omega \right)+2\right)+\omega \right] \sin [\eta_0 (2 k+\omega )]\Big\} \ .
    \label{eq:I2}
\eea
Leaving aside the trigonometric contributions and taking the limit $\Delta \to 0$, so that $\omega \to k$, one recovers once more Maldacena's result.\footnote{The original result in~\cite{maldacena} contained a typo, which was also noted in~\cite{pajer}.}
%%%%%%%%%%%%%%%%%%%%%%%%%%%%%%%%%%
\subsection[Three gravitons]{\sc Three gravitons}\label{sec:threegravitons}
%%%%%%%%%%%%%%%%%%%%%%%%%%%%%%%%%%%%%
Finally, the cubic action for tensor perturbations in the uniform-density gauge is~\cite{maldacena,css}
\beq
    \mathcal{S}_{\gamma \gamma \gamma}^{(3)}=\frac{M_p^2}{8}\int d^4 x \, e^{a/3} \, \left[2 \, \gamma_{ik} \, \gamma_{jl} \, - \, \gamma_{ij} \, \gamma_{kl}\right] \, \partial_k \partial_l \, \gamma_{ij} \ .
    \label{eq:cubact3}
\eeq
Since no factors of $\epsilon$ appear in the interaction, the absence of turning-point contributions is manifest in this case. The algebra involving the polarization tensors is identical to what can be found in~\cite{maldacena}, and consequently the three-point function is
\beq
    \langle \gamma^{s_1}_{\mathbf{k}_1} 
    \gamma^{s_2}_{\mathbf{k}_2} 
    \gamma^{s_3}_{\mathbf{k}_3} \rangle 
    = (2\pi)^3 \, \delta^{(3)}\left(\sum_i \mathbf{k}_i\right)
    \, \frac{\mathcal{H}^4}{M_{pl}^4}
    \frac{1}{8 \, k_1^3 \, k_2^3 \, k_3^3} \, (-4)
    \left(
    \epsilon^{s_1}_{ii'} \,
    \epsilon^{s_2}_{jj'} \,
    \epsilon^{s_3}_{ll'} \,
    t_{ijl} \,
    t_{i'j'l'}
    \right)
    \, I_3 \ ,
    \label{eq:amp3}
\eeq
where $t_{ijl}$ is
\begin{equation*}
    t_{ijl} = k_{i}^2\, \delta_{jl} 
    + k_{j}^3\, \delta_{il} 
    + k_{l}^1\, \delta_{ij} \ .
    \label{eq:t}
\end{equation*}
The time integral $I_3$ is
\beq
    I_3 = \mathrm{Re} \left[
    - i \int_{\eta_0}^{0} \frac{d\eta}{\eta^{2}}
    (1 - i k_1 \eta)(1 - i k_2 \eta)(1 - i k_3 \eta)
    e^{i K \eta}
    \right] \ .
    \label{eq:I3}
\eeq
with $K \equiv k_1 + k_2 + k_3$, 
and in the equilateral configuration one finally obtains
\beq
    I_3=\left(\frac{1}{\eta_0}-\frac{\eta_0 \, k^2}{3}\right) \sin (3 \, \eta_0 \, k)-\frac{10}{9} \, k \, \cos (3 \, \eta_0 \, k)-\frac{17}{9}k \ . 
    \label{eq:I3res}
\eeq
Once more, ignoring the trigonometric terms, one recovers Maldacena's equilateral result.

As we have already emphasized, the only novelty present in all these amplitudes involving tensor modes are the oscillatory features induced by the underlying bounce scenario, so that the absence of turning-point contributions increases the dominance of scalar non-Gaussian signals over their tensor counterparts. Moreover, the absence of $\Delta$ in these expressions implies that the oscillatory behavior of the three-graviton signal is perfectly harmonic in $k$, in contrast to what happens in the other amplitudes that also involve scalar perturbations. If these effects were to prove observable in future experiments, this distinction might provide a handle for an independent determination of $\Delta$. 

%%%%%%%%%%%%%%%%%%%%%%%%%%%%%%
\section*{\sc Acknowledgments}
%%%%%%%%%%%%%%%%%%%%%%%%%%%%%%
\vskip 12pt
I am very grateful to A.~Sagnotti for collaboration at the early stages of this project and to A.~Gruppuso for useful comments. This work was supported in part by Scuola Normale and in part by INFN (IS GSS-Pi).


\newpage
\begin{thebibliography}{99}

\bibitem{stringtheory}
For reviews see: \\
M.~B.~Green, J.~H.~Schwarz and E.~Witten, ``Superstring Theory'', 2 vols., Cambridge Univ. Press (1987); \\
J.~Polchinski, ``String theory'', 2 vols. Cambridge, UK: Cambridge Univ. Press (1998);  \\
C.~V.~Johnson, ``D-branes,'' Cambridge Univ. Press (2003); \\
B.~Zwiebach, ``A first course in string theory,'' Cambridge Univ. Press (2004); \\
K.~Becker, M.~Becker and J.~H.~Schwarz,
``String theory and M-theory: A modern introduction'' Cambridge, UK: Cambridge Univ.
Press (2007); \\
E.~Kiritsis, ``String theory in a nutshell,'' Princeton Univ. Press (2007);\\
P.~West, ``Introduction to strings and branes,'' Cambridge Univ. Press (2012).
  %%CITATION = INSPIRE-1190041;%%

\bibitem{ssb_review} 
For a recent review, see: \\
E.~Dudas, J.~Mourad and A.~Sagnotti,
%``Supersymmetry Breaking with Fields, Strings and Branes,''
[arXiv:2511.04367 [hep-th]].

\bibitem{climbing}
E.~Dudas, N.~Kitazawa and A.~Sagnotti,
%``On Climbing Scalars in String Theory,''
Phys. Lett. B \textbf{694} (2011), 80
%doi:10.1016/j.physletb.2010.09.040
[arXiv:1009.0874 [hep-th]].

\bibitem{planck_lack}
P.~A.~R.~Ade \textit{et al.} [Planck],
%``Planck 2013 results. XXIII. Isotropy and statistics of the CMB,''
Astron. Astrophys. \textbf{571} (2014), A23
%doi:10.1051/0004-6361/201321534
[arXiv:1303.5083 [astro-ph.CO]];
P.~A.~R.~Ade \textit{et al.} [Planck],
%``Planck 2015 results. XVI. Isotropy and statistics of the CMB,''
Astron. Astrophys. \textbf{594} (2016), A16
%doi:10.1051/0004-6361/201526681
[arXiv:1506.07135 [astro-ph.CO]];
Y.~Akrami \textit{et al.} [Planck],
%``Planck 2018 results. VII. Isotropy and Statistics of the CMB,''
Astron. Astrophys. \textbf{641} (2020), A7
%doi:10.1051/0004-6361/201935201
[arXiv:1906.02552 [astro-ph.CO]].

\bibitem{orientifolds}
A.~Sagnotti, 
%``Open Strings And Their Symmetry Groups,'' 
in Cargese '87, ``Non-Perturbative Quantum Field
Theory'', eds. G. Mack et al (Pergamon Press, 1988), p. 521,
arXiv:hep-th/0208020;
%%CITATION = HEP-TH 0208020;%%
G.~Pradisi and A.~Sagnotti,
%``Open String Orbifolds,''
Phys.\ Lett.\ {\bf B 216} (1989) 59;
%%CITATION = PHLTA,B216,59;%%
P.~Horava,
%``Strings On World Sheet Orbifolds,''
Nucl.\ Phys.\ {\bf B 327} (1989) 461;
%%CITATION = NUPHA,B327,461;%%
P.~Horava, 
%``Background Duality Of Open String Models,''
Phys.\ Lett.\ {\bf B 231} (1989) 251;
%%CITATION = PHLTA,B231,251;%%
M.~Bianchi and A.~Sagnotti,
%``On The Systematics Of Open String Theories,''
Phys.\ Lett.\ {\bf B 247} (1990) 517;
%%CITATION = PHLTA,B247,517;%%
M.~Bianchi and A.~Sagnotti,
%``Twist Symmetry And Open String Wilson Lines,''
Nucl.\ Phys.\ {\bf B 361} (1991) 519;
%%CITATION = NUPHA,B361,519;%%
M.~Bianchi, G.~Pradisi and A.~Sagnotti,
%``Toroidal compactification and symmetry breaking in open string theories,''
Nucl.\ Phys.\ {\bf B 376} (1992) 365;
%%CITATION = NUPHA,B376,365;%%
A.~Sagnotti,
 %``A Note on the Green-Schwarz mechanism in open string theories,''
 Phys.\ Lett.\  {\bf B 294} (1992) 196
 [arXiv:hep-th/9210127].\\
 %%CITATION = PHLTA,B294,196;%%
 For reviews, see:
E.~Dudas,
%``Theory and phenomenology of type I strings and M-theory,''
Class.\ Quant.\ Grav.\  {\bf 17} (2000) R41 [arXiv:hep-ph/0006190];
%%CITATION = HEP-PH 0006190;%%
C.~Angelantonj and A.~Sagnotti,
%``Open strings,''
Phys.\ Rept.\  {\bf 371} (2002) 1 [Erratum-ibid.\  {\bf 376} (2003)
339] [arXiv:hep-th/0204089];
%%CITATION = HEP-TH 0204089;%%
C.~Angelantonj and I.~Florakis,
%``A Lightning Introduction to String Theory,''
%doi:10.1007/978-981-19-3079-9\_53-1
[arXiv:2406.09508 [hep-th]].


\bibitem{susy95}
A.~Sagnotti,
%``Some properties of open string theories,''
[arXiv:hep-th/9509080 [hep-th]];
%179 citations counted in INSPIRE as of 31 May 2021
A.~Sagnotti,
%``Surprises in open string perturbation theory,''
Nucl. Phys. B Proc. Suppl. \textbf{56} (1997), 332
%doi:10.1016/S0920-5632(97)00344-7
[arXiv:hep-th/9702093 [hep-th]].
%157 citations counted in INSPIRE as of 31 May 2021

\bibitem{usp32}
S.~Sugimoto,
%``Anomaly cancellations in type I D9-D9-bar system and the USp(32)  string
%theory,''
Prog.\ Theor.\ Phys.\  {\bf 102} (1999) 685 [arXiv:hep-th/9905159].
%%CITATION = HEP-TH 9905159;%%

\bibitem{bsb_rev}
J.~Mourad and A.~Sagnotti,
%``An Update on Brane Supersymmetry Breaking,''
[arXiv:1711.11494 [hep-th]].
%59 citations counted in INSPIRE as of 25 Aug 2025


\bibitem{so1616} 
L.~J.~Dixon and J.~A.~Harvey,
%``String Theories in Ten-Dimensions Without Space-Time Supersymmetry,''
Nucl. Phys. B \textbf{274} (1986), 93.
%doi:10.1016/0550-3213(86)90619-X
%390 citations counted in INSPIRE as of 01 Jun 2021
L.~Alvarez-Gaume, P.~H.~Ginsparg, G.~W.~Moore and C.~Vafa,
%``An O(16) x O(16) Heterotic String,''
Phys. Lett. B \textbf{171} (1986), 155.
%doi:10.1016/0370-2693(86)91524-8
%313 citations counted in INSPIRE as of 01 Jun 2021

\bibitem{dm}
E.~Dudas and J.~Mourad,
% ``Brane solutions in strings with broken supersymmetry and dilaton tadpoles,''
 Phys.\ Lett.\  {\bf B 486} (2000) 172
 [arXiv:hep-th/0004165].
 %%CITATION = PHLTA,B486,172;%%

\bibitem{russo}
J.~G.~Russo,
 % ``Exact solution of scalar-tensor cosmology with exponential potentials and transient acceleration,''
  Phys.\ Lett.\  {\bf B 600} (2004), 185
  [arXiv:hep-th/0403010].
  %%CITATION = PHLTA,B600,185;%%

 \bibitem{inflation}
 A.~A.~Starobinsky,
%  ``A New Type of Isotropic Cosmological Models Without Singularity,''
  Phys.\ Lett.\ {\bf B 91} (1980) 99;
 %%CITATION = PHLTA,B91,99;%%
  D.~Kazanas,
 %``Dynamics of the Universe and Spontaneous Symmetry Breaking,''
  Astrophys.\ J.\  {\bf 241} (1980) L59;
 %%CITATION = ASJOA,241,L59;%%
  K.~Sato,
%``Cosmological Baryon Number Domain Structure and the First Order Phase Transition of a Vacuum,''
  Phys.\ Lett.\  {\bf B 99} (1981) 66;
 %%CITATION = PHLTA,B99,66;%%
 A.~H.~Guth,
% ``The Inflationary Universe: A Possible Solution to the Horizon and Flatness Problems,''
 Phys.\ Rev.\ {\bf D 23} (1981) 347;
  %%CITATION = PHRVA,D23,347;%%
 A.~D.~Linde,
% ``A New Inflationary Universe Scenario: A Possible Solution of the Horizon, Flatness, Homogeneity, Isotropy and Primordial Monopole Problems,''
  Phys.\ Lett.\ {\bf B 108} (1982) 389;
  %%CITATION = PHLTA,B108,389;%%
   A.~Albrecht and P.~J.~Steinhardt,
%``Cosmology for Grand Unified Theories with Radiatively Induced Symmetry Breaking,''
  Phys.\ Rev.\ Lett.\  {\bf 48} (1982) 1220;
   A.~D.~Linde,
%``Chaotic Inflation,''
  Phys.\ Lett.\ {\bf B 129} (1983) 177.
 %%CITATION = PHLTA,B129,177;%%
For reviews, see: 
 N.~Bartolo, E.~Komatsu, S.~Matarrese and A.~Riotto,
%``Non-Gaussianity from inflation: Theory and observations,''
  Phys.\ Rept.\  {\bf 402} (2004) 103
  [astro-ph/0406398].\\
  %%CITATION = ASTRO-PH/0406398;%%
V.~Mukhanov,
  ``Physical foundations of cosmology,''
  Cambridge Univ. Press (2005); \\
S.~Weinberg, 
``Cosmology,''
 Oxford Univ. Press (2008); \\
D.~H.~Lyth and A.~R.~Liddle,
``The primordial density perturbation: Cosmology, inflation and the origin of structure,''
  Cambridge Univ. Press (2009); \\
  D.~S.~Gorbunov and V.~A.~Rubakov,
  ``Introduction to the theory of the early universe: Cosmological perturbations and inflationary theory,''
  World Scientific (2011);\\
  %doi:10.1142/7874
  %%CITATION = doi:10.1142/7874;%%
  %15 citations counted in INSPIRE as of 04 Aug 2017
 J.~Martin, C.~Ringeval and V.~Vennin,
 % ``Encyclopaedia Inflationaris,''
  Phys.\ Dark Univ.\  {\bf 5-6} (2014) 75
  [arXiv:1303.3787 [astro-ph.CO]];\\
  %%CITATION = doi:10.1016/j.dark.2014.01.003;%%
    N.~Vittorio, ``Cosmology,''
 CRC Press (2018);\\
  D.~Baumann, ``Cosmology,''
 Cambridge Univ. Press (2022).

\bibitem{dkps}
E.~Dudas, N.~Kitazawa, S.~P.~Patil and A.~Sagnotti,
%``CMB Imprints of a Pre-Inflationary Climbing Phase,''
JCAP \textbf{05} (2012), 012
%doi:10.1088/1475-7516/2012/05/012
[arXiv:1202.6630 [hep-th]];
N.~Kitazawa and A.~Sagnotti,
%``Pre-inflationary clues from String Theory?,''
JCAP \textbf{04} (2014), 017
%doi:10.1088/1475-7516/2014/04/017
[arXiv:1402.1418 [hep-th]].

\bibitem{cm}
V.~F.~Mukhanov and G.~V.~Chibisov,
% ``Quantum Fluctuations and a Nonsingular Universe,''
JETP Lett. \textbf{33} (1981), 532.
%1682 citations counted in INSPIRE as of 03 Jun 2021
For a review, see:
V.~F.~Mukhanov, H.~A.~Feldman and R.~H.~Brandenberger,
%``Theory of cosmological perturbations. Part 1. Classical perturbations. Part 2. Quantum theory of perturbations. Part 3. Extensions,''
Phys. Rept. \textbf{215} (1992), 203.
%doi:10.1016/0370-1573(92)90044-Z
%3462 citations counted in INSPIRE as of 27 Nov 2023

\bibitem{gkmns}
A.~Gruppuso and A.~Sagnotti,
%``Observational Hints of a Pre--Inflationary Scale?,''
Int. J. Mod. Phys. D \textbf{24} (2015) no.12, 1544008
%doi:10.1142/S0218271815440083
[arXiv:1506.08093 [astro-ph.CO]];
%31 citations counted in INSPIRE as of 11 Sep 2025
A.~Gruppuso, N.~Kitazawa, N.~Mandolesi, P.~Natoli and A.~Sagnotti,
%``Pre-Inflationary Relics in the CMB?,''
Phys. Dark Univ. \textbf{11} (2016), 68
%doi:10.1016/j.dark.2015.12.001
[arXiv:1508.00411 [astro-ph.CO]];
A.~Gruppuso, N.~Kitazawa, M.~Lattanzi, N.~Mandolesi, P.~Natoli and A.~Sagnotti,
%``The Evens and Odds of CMB Anomalies,''
Phys. Dark Univ. \textbf{20} (2018), 49
%doi:10.1016/j.dark.2018.03.002
[arXiv:1712.03288 [astro-ph.CO]].

\bibitem{ms}
M.~Meo and A.~Sagnotti,
%``On Pre-Inflationary non Gaussianities,''
JHEP \textbf{12} (2025), 016
%doi:10.1007/JHEP12(2025)016
[arXiv:2510.01360 [hep-th]];
M.~Meo, Master Thesis [arXiv:2510.23377 [hep-th]].

\bibitem{gasp_ven}
M.~Gasperini and G.~Veneziano,
%``Pre - big bang in string cosmology,''
Astropart. Phys. \textbf{1} (1993), 317-339
%doi:10.1016/0927-6505(93)90017-8
[arXiv:hep-th/9211021 [hep-th]];
%1017 citations counted in INSPIRE as of 09 Sep 2025
M.~Gasperini and G.~Veneziano,
%``The Pre - big bang scenario in string cosmology,''
Phys. Rept. \textbf{373} (2003), 1-212
%doi:10.1016/S0370-1573(02)00389-7
[arXiv:hep-th/0207130 [hep-th]].
%818 citations counted in INSPIRE as of 09 Sep 2025


\bibitem{Planck}
P.A.R.~Ade \textit{et al.} [Planck],
%``Planck 2013 Results. XXIV. Constraints on primordial non-Gaussianity,''
Astron. Astrophys. \textbf{571} (2014), A24
%doi:10.1051/0004-6361/201321554
[arXiv:1303.5084 [astro-ph.CO]];
%840 citations counted in INSPIRE as of 15 Sep 2025
P.~A.~R.~Ade \textit{et al.} [Planck],
%``Planck 2015 results. XVII. Constraints on primordial non-Gaussianity,''
Astron. Astrophys. \textbf{594} (2016), A17
%doi:10.1051/0004-6361/201525836
[arXiv:1502.01592 [astro-ph.CO]].
%862 citations counted in INSPIRE as of 15 Sep 2025
Y.~Akrami \textit{et al.} [Planck],
%``Planck 2018 results. IX. Constraints on primordial non-Gaussianity,''
Astron. Astrophys. \textbf{641} (2020), A9
%doi:10.1051/0004-6361/201935891
[arXiv:1905.05697 [astro-ph.CO]].
%904 citations counted in INSPIRE as of 15 Sep 2025

\bibitem{forecasts}
M.~Abitbol \textit{et al.} [Simons Observatory],
%``The Simons Observatory: science goals and forecasts for the enhanced Large Aperture Telescope,''
JCAP \textbf{08} (2025), 034
%doi:10.1088/1475-7516/2025/08/034
[arXiv:2503.00636 [astro-ph.IM]];
%38 citations counted in INSPIRE as of 15 Sep 2025
W.~Sohn and J.~Fergusson,
%``CMB-S4 forecast on the primordial non-Gaussianity parameter of feature models,''
Phys. Rev. D \textbf{100} (2019) no.6, 063536
%doi:10.1103/PhysRevD.100.063536
[arXiv:1902.01142 [astro-ph.CO]].
%15 citations counted in INSPIRE as of 15 Sep 2025

\bibitem{ss} J.~Scherk and J.~H.~Schwarz,
%``How to Get Masses from Extra Dimensions,''
Nucl. Phys. B \textbf{153} (1979), 61.
%doi:10.1016/0550-3213(79)90592-3;
%1265 citations counted in INSPIRE as of 16 Nov 2023

\bibitem{maldacena}
J.~M.~Maldacena,
%``Non-Gaussian features of primordial fluctuations in single field inflationary models,''
JHEP \textbf{05} (2003), 013
%doi:10.1088/1126-6708/2003/05/013
[arXiv:astro-ph/0210603 [astro-ph]].
%3023 citations counted in INSPIRE as of 06 Jul 2025
See also:\\
J.~Noller and J.~Magueijo,
%``Non-Gaussianity in single field models without slow-roll,''
Phys. Rev. D \textbf{83} (2011), 103511
%doi:10.1103/PhysRevD.83.103511
[arXiv:1102.0275 [astro-ph.CO]];
%40 citations counted in INSPIRE as of 06 Jul 2025
H.~Collins,
%``Primordial non-Gaussianities from inflation,''
[arXiv:1101.1308 [astro-ph.CO]].
%35 citations counted in INSPIRE as of 06 Jul 2025

\bibitem{lm}
 F.~Lucchin and S.~Matarrese,
%``Power Law Inflation,''
Phys. Rev. D \textbf{32} (1985), 1316.
% doi:10.1103/PhysRevD.32.1316
%944 citations counted in INSPIRE as of 15 Feb 2024

\bibitem{pajer}
E.~Pajer, 
%``Building a Boostless Bootstrap for the Bispectrum,''
JCAP \textbf{01} (2021), 023
%doi:10.1088/1475-7516/2021/01/023
[arXiv:2010.12818 [hep-th]];
P.~McFadden and K.~Skenderis
%``Cosmological 3-point correlators from holography,''
JCAP \textbf{06} (2011), 030
[arXiv:1104.3894 [hep-th]].

\bibitem{css}
G.~Cabass, D.~Stefanyszyn, J.~Supel and A.~Thavanesan,
%``On Graviton non-Gaussianities in the Effective Field Theory of Inflation,''
JHEP \textbf{10} (2022), 154 
[arXiv:2209.00677 [hep-th]].

\end{thebibliography}
\end{document}